\documentclass[prl,twocolumn,superscriptaddress]{revtex4}
\usepackage{graphicx}
\usepackage{epstopdf}
\usepackage{epsfig}
\usepackage{times}
\usepackage{amsmath}
\usepackage{color}

\begin{document}

\title{Sustained quantum coherence and entanglement in the avian compass}

\author{Erik M. Gauger$^*$}
\affiliation{Department of Materials, University of Oxford, Parks Rd, Oxford OX1 3PH, UK}

\author{Elisabeth Rieper\footnote{These authors have contributed equally to the work reported here.}}
\affiliation{Centre for Quantum Technologies, National University of Singapore, Singapore}

\author{John J. L. Morton}
\affiliation{Department of Materials, University of Oxford, Parks Rd, Oxford OX1 3PH, UK}
\affiliation{Clarendon Laboratory, University of Oxford, Parks Rd, OX1 3PU, UK}

\author{Simon C. Benjamin}
\email{s.benjamin@qubit.org}
\affiliation{Centre for Quantum Technologies, National University of Singapore, Singapore}
\affiliation{Department of Materials, University of Oxford, Parks Rd, Oxford OX1 3PH, UK}

\author{Vlatko Vedral}
\affiliation{Centre for Quantum Technologies, National University of Singapore, Singapore}
\affiliation{Clarendon Laboratory, University of Oxford, Parks Rd, OX1 3PU, UK}
\affiliation{Department of Physics, National University of Singapore, Singapore}

\date{\today}

\begin{abstract}In artificial systems, quantum superposition and entanglement typically decay rapidly unless cryogenic temperatures are used. Could life have evolved to exploit such delicate phenomena? Certain migratory birds have the ability to sense very subtle variations in the Earth's magnetic field. Here we apply quantum information theory and the widely accepted `radical pair' model to analyse recent experimental observations of the avian compass. We find that superposition and entanglement are sustained in this living system for at least tens of microseconds, exceeding the durations achieved in the best comparable man-made molecular systems. This conclusion is starkly at variance with the view that life is too `warm and wet' for such quantum phenomena to endure. 
\end{abstract}

\maketitle

Recently several authors have raised the intriguing possibility that living systems may use non-trivial quantum effects to optimise some tasks.  Studies range from the role of quantum physics in photosynthesis~\cite{engel07,mohseni08,plenio08,sarovar09} and in natural selection itself~\cite{lloyd09}, through to the observation that `warm and wet' living systems can embody entanglement given a suitable cyclic driving~\cite{cai08}. In this Letter, we examine quantum phenomena in the process of magnetoreception -- the ability to sense characteristics of the surrounding magnetic field. 

There are a several mechanisms by which this sense may operate~\cite{johnsen08}. In certain species (including certain birds~\cite{ritz00,zebra},  fruit flies~\cite{gegear08,yoshii09} and even plants~\cite{ahmad07}), the evidence supports a so-called Radical Pair (RP) mechanism. This process involves the quantum evolution of a spatially-separated pair of electron spins~\cite{ritz00,ritz04}, and such a model is supported by several results from the field of spin chemistry~\cite{timmel04, liu05, miura06, rodgers09, rodgers09a}. An artificial chemical compass operating according to this principle has been demonstrated experimentally~\cite{maeda08}, and a very recent theoretical study examines the presence of entanglement within such a system~\cite{cai09}. Here, we  consider the timescales for the persistence of full quantum coherence, and entanglement, within a specific living system: the European Robin. Our analysis uses recent data from experiments on live birds. We conclude that the RP model implies a decoherence time in the birds' compass which is extraordinarily long -- beyond that of any artificial molecular system.

By manipulating a captive bird's magnetic environment and recording its response, one can make inferences about the mechanism of the magnetic sensor~\cite{wiltschko72, wiltschko02,ritz04,ritz09}. Specifically, European Robins are only sensitive to the inclination and not the polarization of the magnetic field \cite{wiltschko72}, and this sensor is evidently activated by photons entering the bird's eye~\cite{wiltschko02,zapka09}. Importantly for the present analysis, a very small oscillating magnetic field can disrupt the bird's ability to orientate~\cite{ritz04, ritz09}. It is also significant that birds are able to `train' to different field strengths, suggesting that the navigation sense is robust, and unlikely to depend on very special values for the parameter in the model~\cite{ritz09}.\\

\begin{figure}[b]\centering
\includegraphics[width=7cm]{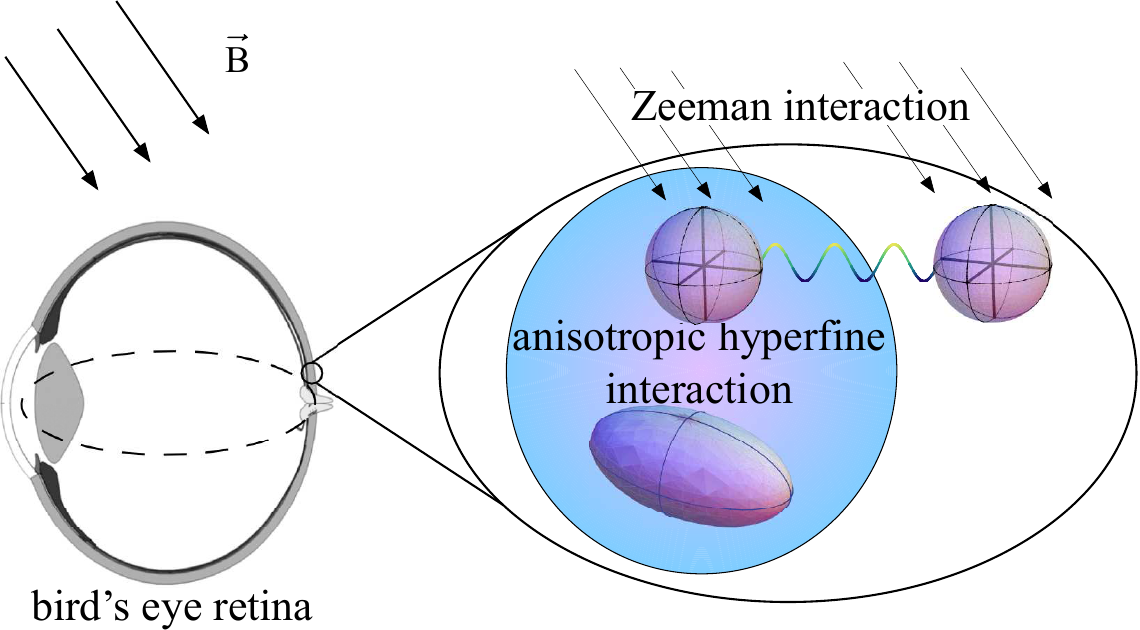}\\
\caption{ \label{eye} According to the RP model, the back of the bird's eye contains numerous molecules for magnetoreception~\cite{solovyov09}. These molecules give rise to a pattern, discernible to the bird, which indicates the orientation of the field. Note that this implies that the molecules involved are at least fixed in orientation, and possibly ordered with respect to one another~\cite{ritz00}. In the simplest variant, each such molecule involves three crucial components (see inset): there are two electrons, initially photo-excited to a singlet state, and a nuclear spin that couples to {\em one} of the electrons. This coupling is anisotropic, so that the molecule has a directionality to it. }  
\end{figure}

The basic idea of the RP model is as follows: there are molecular structures in the bird's eye which can each absorb an optical photon and give rise to a spatially separated electron pair in a singlet spin state. Because of the differing local environments of the two electron spins, a singlet-triplet evolution occurs. This evolution depends on the inclination of the molecule with respect to the Earth's magnetic field. Recombination occurs either from the singlet or triplet state, leading to different chemical end products. The concentration of these products constitutes a chemical signal correlated to the Earth's field orientation. The specific molecule involved is unknown. 

Making as few assumptions as possible about the detailed structure of the molecule, we examine a family of models with the necessary complexity to support this RP mechanism. Our aim is to understand whether full quantum coherence and entanglement exist for long durations in the European Robin's compass system.  
Figure~\ref{eye} depicts the most basic form of the model: two electronic spins~\cite{ritz00} and one nuclear spin. The nucleus interacts with only one of the electron spins, thus providing the asymmetry required for singlet-triplet oscillations. In this model, as with the other models we consider, we employ the Hamiltonian corresponding to the system once the two electrons have become separated. That is, our $t=0$ corresponds to the moment of RP formation. 

The anisotropic hyperfine tensor coupling the nucleus and electron $1$, is conveniently written in its diagonal basis $A={\rm diag}(A_x, A_y, A_z)$, and we assume an axially symmetric (or cigar-shaped) molecule with $A_z = 10^{-5} ~ \rm{meV}$ and $A_x = A_y = A_z /2$. This is the simplest assumption that can provide us with directionality, and we have chosen the general shape and magnitude of the tensor to be consistent with \cite{efimova09}. The Hamiltonian is
\begin{equation}
H =  \hat{I} \cdot {\bf A} \cdot \hat{S}_1 + \gamma {\bf B} \cdot (\hat{S}_1 + \hat{S}_2),
\nonumber
\end{equation}
where $\hat{I}$ is the nuclear spin operator, $\hat{S}_i=(\sigma_x,\sigma_y,\sigma_z)_i$ are the electron spin operators ($i=1,2$), $\bf{B}$ is the magnetic field vector and $\gamma =  \frac{1}{2} \mu_0 g $  the gyromagnetic ratio with $\mu_0$ being Bohr's magneton and $g=2$ the g-factor. The factor 1/2 in the gyromagnetic ratio accounts for the fact that we have a spin one-half system, but we will use Pauli matrices such as $\sigma_z={\rm diag}\{1,-1\}$ etc. 
Here only one electron is coupled to one nucleus, whereas the remote electron is so weakly interacting that we describe it as free. 

We have also considered a family of variants involving different hyperfine tensors, adding a second nuclear spin (following previous studies where more than one nucleus couples to the system~\cite{ritz09,rodgers09a, rodgers07, cai09}), and replacing the nuclear asymmetry with an anisotropic electron g-factor. These models, and the results of the corresponding simulations, are presented in the Supplementary Material~\cite{SupMat}. In essence all models give rise to the same qualitative behavior as the basic model described here. This is not surprising since there is a basic underlying principle: The electron spins of the RP must be protected from an irreversible loss of quantum coherence in order to be susceptible to the experimentally applied RF field. The extremely low strength of this applied field dictates the timescale over which quantum coherence must be preserved. Thus the inference of extraordinarily long coherence times does not vary significantly over the various models.

Generally, the magnetic field we employ is
\begin{eqnarray}
\bf{B} &  = & B_0 (\cos \varphi \sin \vartheta, \sin \varphi \sin \vartheta, \cos \vartheta) \nonumber \\
& + & B_{\rm rf} \cos \omega t (\cos \phi  \sin \theta, \sin \phi \sin \theta, \cos \theta),
\label{BfieldEqn}
\end{eqnarray}
where $B_0 = 47~{\rm \mu}$T is the Earth's magnetic field in Frankfurt \cite{ritz09}, and the angles describe the orientation of magnetic field to the basis of the HF tensor. $B_{\rm rf} = 150$~nT is an additional oscillatory field only applied in our simulations where explicitly mentioned. For resonant excitation with the uncoupled electron spin, $\hbar \omega = 2 \gamma B_0$, so that $\nu = \omega / (2 \pi) = 1.316~\rm{MHz}$.

The axial symmetry of the HF tensor allows us to set $\varphi = 0$ and focus on $\vartheta$ in the range $[0, \pi/2]$ without loss of generality. For the oscillatory field we set $\phi = 0$. 

To model the dynamics of the system with a quantum master equation (ME) approach, we add two `shelving states' to the 8 dimensional Hilbert space of the three spins. We employ operators to represent the spin-selective relaxation into the singlet shelf $\vert  S \rangle$ from the electron singlet state, or the triplet shelf $\vert T \rangle$ from the triplet configurations. One of the two events will occur, and the final populations of  $\vert  S \rangle$ and  $\vert  T \rangle$ give  the singlet and triplet yield.

With the usual definiton of singlet $\vert  s \rangle$ and triplet states $\vert  t_i  \rangle$ in the electronic subspace, while $\vert  \uparrow \rangle$ and $\vert  \downarrow \rangle$ describe the states of the nuclear spin, we define the following decay operators: 
$ P_{S, \uparrow}  = \vert S \rangle \langle s, \uparrow \vert$, 
$P_{T_0, \uparrow}  = \vert T \rangle \langle t_0, \uparrow \vert$,
$P_{T_+, \uparrow}  = \vert T \rangle \langle t_+, \uparrow \vert$,
$P_{T_-, \uparrow}  = \vert T \rangle \langle t _-, \uparrow \vert$, 
and similarly for the `down' nuclear states. This gives a total of two singlet and six triplet projectors to discriminate the respective decays with a standard Lindblad ME,
\begin{eqnarray}
\dot{\rho} =  -\frac{i}{\hbar} [H, \rho ]  +  k  \sum_{i=1}^8 P_i \rho P_i^{\dagger}   - \frac{1}{2}  P_i^{\dagger} P_i \rho + \rho  P_i^{\dagger} P_{i} ~. 
\label{masterEqn}
\end{eqnarray}
For simplicity and because this choice corresponds exactly to the expression for the singlet yield used in the previous literature, all eight projectors have been assigned the same decay rate $k$. Note that  Eqn. (\ref{masterEqn}) does not yet contain environmental noise, though this will not alter our estimate of $k$ (see Supplementary Material~\cite{SupMat}).

\begin{figure}[h!]
\begin{center}
\includegraphics[width=0.9\columnwidth]{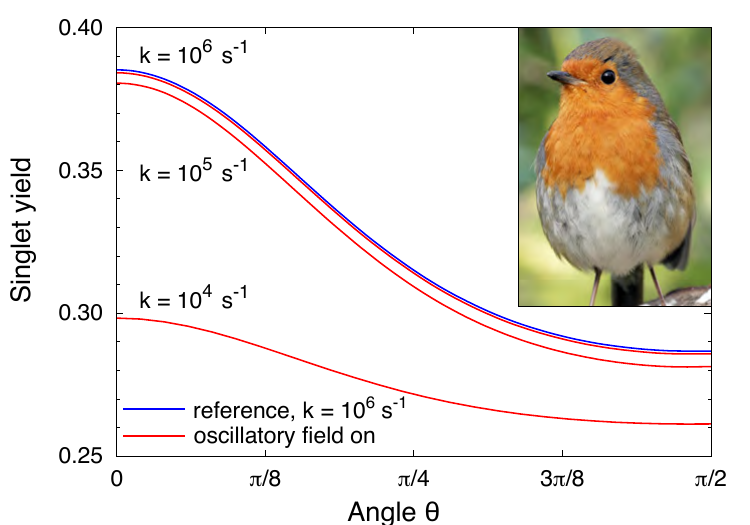} 
\caption{Angular dependence of the singlet yield in the presence of an oscillatory field. The blue curve provides a reference of the singlet yield in the Earth's magnetic field ($B_0 = 47 \rm{\mu T})$. The reference is independent of the decay rate for $k \leq 10^7$~s$^{-1}$, but has been shifted upwards by 0.001 for better visibility. The red curves show the singlet yield when a $150~\mathrm{nT}$ field oscillating at 1.316 MHz (i.e.~resonant with the Zeeman frequency of the uncoupled electron) is superimposed perpendicular to the direction of the static field. This only has an appreciable effect on the singlet yield once $k$ is of order $10^4$~s$^{-1}$. Inset: a European Robin (\copyright~David Jordan).}  
\label{fig:classicalField}
\end{center}
\end{figure}

\begin{figure}
\begin{center}
\includegraphics[width=0.9\columnwidth]{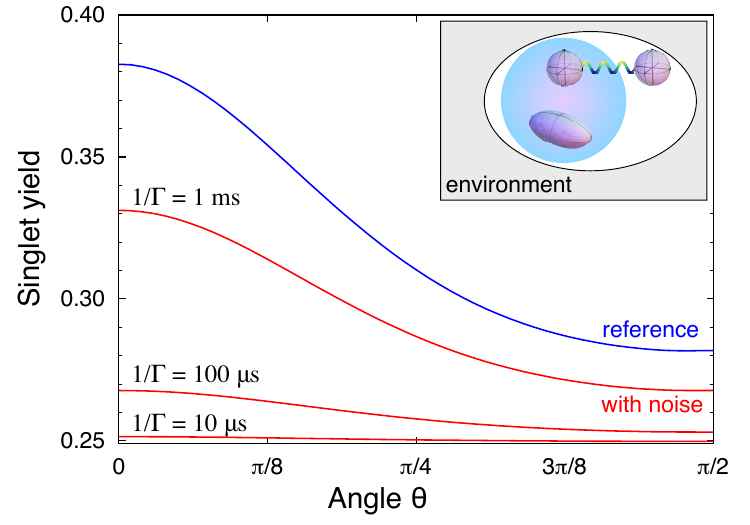} 
\caption{Angular dependence of the singlet yield in the presence of noise (for $k=10^4$). The blue curve provides a reference in the absence of noise and the red curves show the singlet yield for different noise rates. As is apparent from the plot a noise rate $\Gamma > 0.1k$ has a dramatic effect on the magnitude and contrast of the singlet yield. Inset: partitioning between compass and environment.} 
\label{fig:noise}
\end{center}
\end{figure}

The previous literature has been common to employed a Liouville equation to model the RP dynamics. In fact, a term-by-term comparison of the evolution of the density matrix readily confirms that this former approach and our ME are exactly equivalent in the absence of environmental noise. For equal singlet and triplet reaction rates, both give rise to the same singlet yield that is often defined as the integral $\Phi=\int_0^{\infty}\langle{\psi^-}|Tr_n(\rho (t))| \psi^-\rangle k e^{-kt} dt$ in the prior literature. Specifically, the ultimate population of our singlet shelf $| S \rangle$ corresponds to $\Phi$. However, when we presently wish to introduce various kinds of noise operators, our ME approach provides the more intuitive framework.

The initial state of our model $\rho_0$ assigns a pure singlet state to the electrons, and a completely mixed state to the nucleus, 
$\rho(0) = \vert s, \downarrow \rangle \langle s,  \downarrow \vert +  \vert s, \uparrow \rangle \langle s,  \uparrow \vert$ .

We now determine an appropriate choice for our parameter $k$ in Eqn.~\ref{masterEqn}. In Ref.~\onlinecite{ritz09}, the authors report that a perturbing magnetic field of frequency of 1.316~MHz (i.e. the resonance frequency of the `remote' electron) can disrupt the avian compass. They note that this immediately implies a bound on the decay rate $k$ (since the field would appear static for sufficiently rapid decay). Here we aim to refine this bound on $k$ by considering the oscillating magnetic field {\em strength} which suffices to completely disorient the bird's compass, i.e. $150~{\rm nT}$. (Indeed, even a $15~{\rm nT}$ field was reported as being disruptive, but to be conservative in our conclusions we take the larger value here.) To model this effect, we activate the oscillatory field component defined in Eqn.~\ref{BfieldEqn} and examine the singlet yield as a function of the angle between the Earth's field and the molecular axis. Consistent with the experimental work, we find that there is no effect at such weak fields when the oscillatory field is parallel to the Earth's field. Therefore  we set the oscillatory field to be perpendicular. The results are shown in Figure~\ref{fig:classicalField}. We conclude that if the oscillating field is to disorient the bird, as experiments showed, then the decay rate $k$ should be approximately $10^4$ s$^{-1}$ or less. For higher values of $k$ (shorter timescales for the overall process) there is no time for the weak oscillatory field to significantly perturb the system; it relaxes before it has suffered any effect. Such a value for the decay rate is consistent with the long RP lifetimes in certain candidate cryptochrome molecules found in migratory birds~\cite{liedvogel07}. 

Taking the value $k=10^4$ s$^{-1}$, we are able to move to the primary question of interest: how robust this mechanism is against environmental noise. 
There are several reasons for decoherence, e.g.~dipole interactions, electron-electron distance fluctuations and other particles' spin interactions with the electrons. We describe such environmental noise by extending Eqn.~\ref{masterEqn} with a standard Lindblad dissipator~\cite{nielsen00}:
\begin{eqnarray}
\dot{\rho} =  \begin{array}{c}
{\rm RHS\ of}  \\
{\rm Eqn.2}
\end{array} + \sum_i \Gamma_i \left(  L_i \rho L_i^{\dagger}   - \frac{1}{2} \left( L_i^{\dagger} L_i \rho + \rho  L_i^{\dagger} L_i \right) \right)
\end{eqnarray}

This is a general formalism for Markovian noise (the possibility of non-Markovian processes is discussed in the Supplementary Material~\cite{SupMat}). We consider several noise models: first, a physically reasonable generic model in which both phase and amplitude are perturbed with equal probability. In this model, the noise operators $L_i$ are $\sigma_x$, $\sigma_y$, $\sigma_z$ for each electron spin individually (i.e. tensored with identity matrices for the nuclear spin and the other electron spin). This gives a total of six different noise operators $L_i$ and we use the same  decoherence rate $\Gamma$ for all of them. We are now in a position to determine the approximate level of noise which the compass may suffer, by finding the magnitude of $\Gamma$ for which the angular sensitivity fails. This is shown in Fig.~\ref{fig:noise}. Conservatively, we can say that when $\Gamma\geq k$, the angular sensitivity is highly degraded. This is remarkable, since it implies the decoherence time of the two-electron compass system is of order $100~\mu \mathrm{s}$ or more~\footnote{One could assume the bird to be more easily perturbed by the oscillatory field (Fig.~\ref{fig:classicalField}), and obtain a larger $k$. However, that same assumption of high sensitivity should then be applied to the noise analysis (Fig.~\ref{fig:noise}) and in fact the two assumptions would cancel to give the same basic estimate for the decoherence rate. This cancellation is robust, being valid over an order of magnitude in $k$.}! To provide context for this number, we note that the best laboratory experiment involving preservation of a molecular electron spin state has accomplished a decoherence time of $80~\mathrm{\mu s}$~\cite{morton06a}. 

\begin{figure}
\begin{center}
\includegraphics[width=0.9\columnwidth]{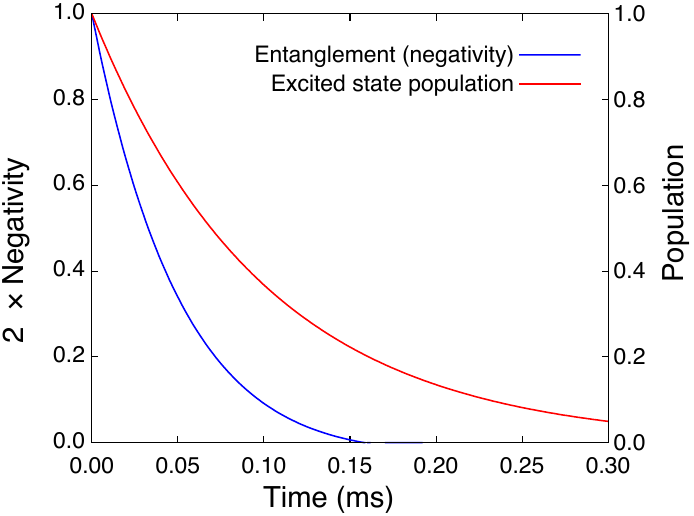} 
\caption{The decline and disappearance of entanglement in the compass system, given the parameter $k$ and the noise severity $\Gamma$ defined above. Here the angle between the Earth's field and the molecular axis is $\pi/4$, although the behavior at other angles is similar.}
\label{fig:negativity}
\end{center}
\end{figure}

In the Supplementary Material~\cite{SupMat} we address the following question: Is there {\em any} noise model, consistent with experimental observations, which would cause rapid decoherence? Initially it seems that such a form of noise exists: the compass mechanism is almost immune to pure phase noise. Even starting from a fully dephased state $(|s\rangle\langle s|+|t_0\rangle\langle t_0|)/2$, the compass operates well. Thus it might seem that strong phase noise could be present, rapidly degrading quantum coherence but permitting the compass to function. However, we show that if such noise were naturally present at the level of $\Gamma_{z}\approx10k$ or higher it would render the bird immune to the weak RF magnetic fields of Ref.~\cite{ritz09}. 

Furthermore, we have performed a systematic study of correlated noise processes. The basic RP model consists of three spins and hence there exist 64 combinations of the form $L = S_i \otimes S_j \otimes S_l$ for $S_{i,j,l} \in \{ I_2, \sigma_x, \sigma_y, \sigma_z$\}. In the Supplementary Material we analyse the combined effect of all these processes added incoherently, as well as a representative subset of individual operators. Strengthening our conclusion from the generic noise model above,  we find that of these models also imply RF field immunity unless coherence is preserved on a timescale that is of order $100~\mu\mathrm{s}$.

It is interesting to characterise the duration of quantum entanglement in this living system. Having inferred approximate values for the key parameters, we can monitor entanglement from the initial singlet generation to the eventual decay. The metric we use is negativity: $N(\rho)=||\rho^{T_A/2}||$, where $||\rho^{T_A}||$ is the trace norm of the partial transpose of the system's density matrix. The transpose is applied to the uncoupled electron, thus performing the natural partitioning between the electron, on one side,  and the coupled electron plus its nucleus, on the other.  Fig.~\ref{fig:negativity} shows how this negativity evolves under our generic noise model. Clearly, the initial singlet state is maximally entangled. Under noise, entanglement falls off at a faster rate than the decay of population from the excited state. 

In summary the reported sensitivity to RF fields implies that both amplitude and phase (and thus entanglement) are indeed protected within the avian compass. The timescales are at least tens of microseconds even for a pure dephasing environment, and hundreds of microseconds for the more general models. It is not clear {\em why} such remarkable protection occurs, but given the widely-accepted RP model together with the recent experimental data~\cite{ritz09}, this conclusion follows.

We thank Earl Campbell, Chris Rodgers, Peter Hore and Kiminori Maeda for stimulating discussions. We thank the National Research Foundation and Ministry of Education of Singapore for support.
EMG acknowledges support from the Marie Curie Early Stage Training network QIPEST (MEST-CT-2005-020505) and the 
QIPIRC (No. GR/S82176/01). 
JJLM and SCB thank the Royal Society for support. JJLM thanks St.\ John's College, Oxford. VV acknowledges financial support from the EPSRC, the Royal Society and the Wolfson Trust in the UK.

\end{document}